\newcounter{myctr}

\documentclass{ws-acs}
\usepackage{url}
\begin{document}

\makeatletter
\def\@biblabel#1{[#1]}
\makeatother

\markboth{Palamara Zlati\'c Scala Caldarelli}{Population Dynamics on Complex Food Webs}

%
\catchline{}{}{}{}{}
%

\title{Population Dynamics on Complex Food Webs}

\author{\footnotesize Gian Marco Palamara\footnote{Microsoft Research Limited and University of Sheffield UK}}

\address{Department of Physics, University of Rome "Sapienza" P.le Aldo Moro 5\\
00185 Rome, Italy\\
gianmarco.palamara@gmail.com}

\author{Vinko Zlati\'c}

\address{Theoretical Physics Division, 
Rudjer Bo\v{s}kovi\'{c} Institute, P.O.Box 180, \\
HR-10002 Zagreb, Croatia\\
vinko@gmail.com}

\address{ISC-CNR, Dep. Physics, University of Rome ``Sapienza'' P.le Moro 5\\
00185 Rome Italy\\}

\author{Antonio Scala}

\address{ISC-CNR, Dep. Physics, University of Rome ``Sapienza'' P.le Moro 5\\
00185 Rome Italy\\
antonio.scala@phys.uniroma1.it}

\author{Guido Caldarelli}

\address{ISC-CNR, Dep. Physics, University of Rome ``Sapienza'' P.le Moro 5\\
00185 Rome Italy\\
Guido.Caldarelli@roma1.infn.it}

\maketitle

\begin{history}
\received{(received date)}
\revised{(revised date)}
\end{history}

\begin{abstract}
In this work we analyse the topological and dynamical properties of a simple model of complex food webs, namely the niche model. In order to underline competition among species, we introduce ``prey" and ``predators" weighted overlap graphs derived from the niche model and compare synthetic food webs with real data. Doing so, we find new tests for the goodness of synthetic food web models and indicate a possible direction of improvement for existing ones. We then exploit the weighted overlap graphs to define a competition kernel for Lotka-Volterra population dynamics and find that for such a model the stability of food webs decreases with its ecological complexity.
\end{abstract}


\section{Introduction}
The study of food webs has attracted the interest of complex systems scientists as one of the clearest example of a network structure whose property can be understood only by looking at the system as a whole. A food web is the collection of the predation relations in an environment and can therefore be naturally described as a network, i.e. a mathematical object composed by vertices (the biological species) and their edges (the predation relations). 

Network structures are ubiquitous and can be found with similar statistical properties in a variety of other situations from WWW \cite{AB99} and the Internet \cite{CMP00} to protein interactions\cite{AJB} and social systems \cite{ASSO}. Despite this similarity, food webs represent  one of the most interesting cases of study thanks to their peculiar topology. In food webs, vertices can be divided in classes thanks to their biological meaning (i.e. prey/predators). Also, the structure is naturally layered when considering the minimum distance of the species from the external resources. All these properties make these structures extremely interesting also as a test-bed for models and algorithms of complex networks. In addition, the application of some ideas developed in the area of computer science points out that biological meaning could hidden in the topology of the food webs \cite{Mercedes}. 

\begin{figure}[th]
\centerline{\psfig{file=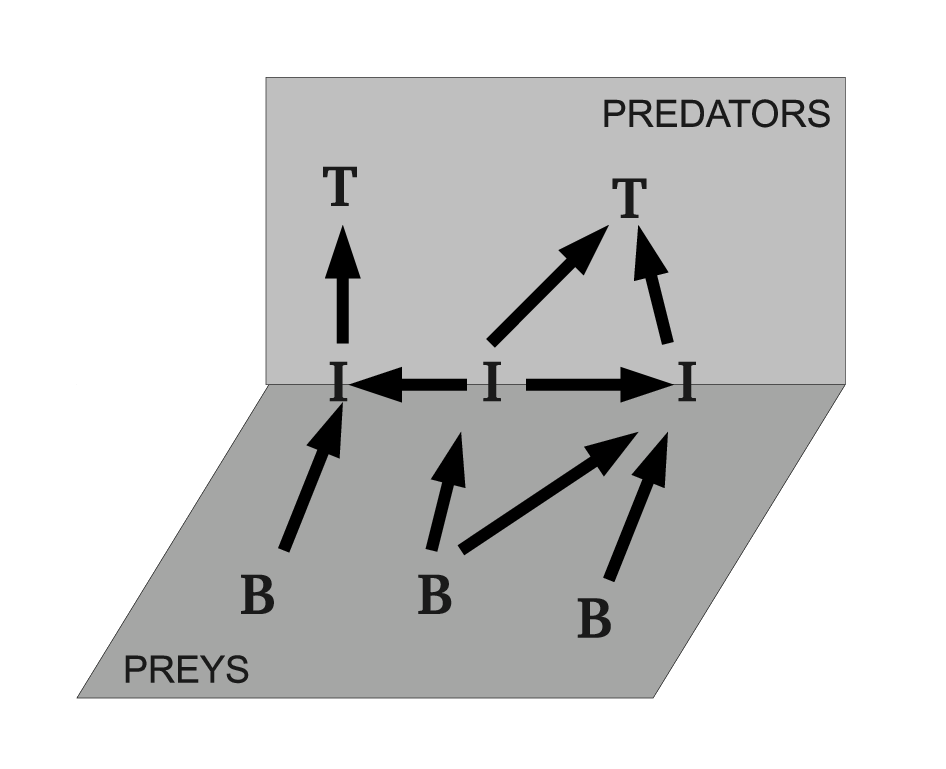,width=10cm}}
\vspace*{8pt}
\caption{Example of a food web. For the sake of clarity, the zeroth node on which basal species feed has been discarded.}\label{fig:PreyPreds}
\end{figure}

Traditionally, the most important quantities in a food web are the number of vertices $N$ and the number of edges $L$. Since the maximum possible number of edges grows as $N^2$ (precisely $N(N-1)$ for a directed graph), it is custom to consider the density of edges $L/N^2$, a quantity also known in ecology as the {\em  directed connectance}. 

In food webs, edges are directed and each of them follows the convention, based on the flux of nutrients, of directing the edge from prey to predator. In this work we will consider only {\em trophic webs}, where all species which have exactly the same predators and prey are merged, i.e. they are represented by a single node. 

Another characterization of food webs is be obtained by considering predation relationships. All the species that have no predators are usually  indicated as the {\em top} (T) species. Similarly, the species with no prey are called {\em basal} species (B). All the others form the {\em intermediate} (I) class.

All the species are ultimately sustained by the transformation into biomass of external resources like water, minerals and sunlight by means of the basal species. It is then customary to describe this situation with the introduction of an external node, called zero-node, which points to all basal species, i.e. to that nodes with only out-edges.  Given this structure, it is easy to define layers of species given by the distance (i.e. the minimum path) towards the zero-node of external resources. Hence, distance is measured as the (minimum) number of edges the biomass has to travel. As happens in the Internet \cite{loops}, some loops accounting for the stability and resilience of these structure \cite{nat} can be present in the system. 

Food webs have been modelled in many ways, and the different models  have been validated with the experimental data available. Here we focus on one recent and successful model on which we will define a topology-determined population dynamics.

\section{Niche Model}
There are many static models of food webs which reproduce the features of real ecosystems such as fractions of top, basal and intermediate species, number of food chains, average chain length, and connectance \cite{connectance}. The simplest way one could think of is to create suitable graphs \cite{bollobas,erdosreny,newmanrand} where (given the linkage density and the number of nodes) directed edges are assigned to randomly chosen pairs of nodes. The agreement between real and such simulated food webs is not very good, as expected. In fact, such a simple model has many unrealistic features such as the assumption that every species can in principle be the predator of every other species.

A first improvement has been the cascade model \cite{cascade} that tries to capture the layered structure of food webs. Williams and Martinez have subsequently improved on the subject by introducing a  static model, called the ``niche model" \cite{niche}, that shows a remarkable agreement between real webs and the synthetic ones generated by the model (particularly true when considering features such as cycles and species similarities).

\begin{figure}[th]
\centerline{\psfig{file=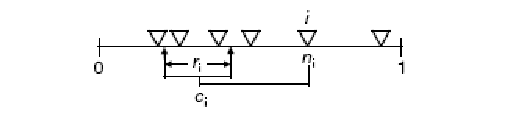,width=10cm}}
\vspace*{8pt}
\caption{Diagram of the niche model. To each of the $S$ species (for example $S=6$, each shown as an inverted triangle) is assigned a ``niche value'' parameter ($n_i$) drawn uniformly from the interval $[0,1]$. Species $i$ consumes
all species falling in a range ($r_i$) that is placed by uniformly drawing the center of the range ($c_i$) from $[r_i/2,n_i]$.}\label{fig:nichemodel}
\end{figure}

The external parameters of the niche model (i.e. the quantities fixed from the beginning) are the number of species $S$ and the directed connectance $C=L/S^2$. To every node is assigned a uniformly distributed number $n_{i}$ into the interval $[0,1]$, the niche space. A species $i$ is characterized by its niche parameter and its list of prey. Prey are chosen for all species according to the following rule (Fig.\ref{fig:nichemodel}): a species $i$ preys on all species $j$ with niche parameters $n_j$ inside a segment of length $r_i$ centred in a position chosen randomly inside the interval $[r_i/2,n_i]$, with $r_i=xn_i$ and $x$ a random variable with probability density function
\begin{equation}
p_x(x)=\beta(x,1,b)=b(1-x)^{(b-1)} \label{beta}
\end{equation}
Choosing $b=(1/2C)-1$ is possible to generate graphs with the desired size and connectance\footnote{In the niche model species with no prey and predators are eliminated and species with the same list of prey and predators, that is trophically identical species, are merged.}.

The niche model estimates the central tendency of empirical data remarkably well \cite{niche}. Its topological and analytical properties have been widely studied \cite{twodegree,analytic} and it has been shown that the predictions of the model are robust with respect to the specific form of the $p_{x}(x)$ chosen\cite{patterns,robustpatterns}.

\section{Projection Graphs}
An aggregated food web with $S$ trophic species can be represented via an $[S \times S]$ adjacency matrix $A$. The elements $a_{ij}$ is taken $1$ if species $j$ preys on species $i$ (directed edge) and $0$ otherwise (no edge). An alternative representation of the ecosystem can be given as a bipartite graph \cite{caldaproj} where two classes of nodes are present: predators (top and intermediate species) and prey (basal and intermediate species) and each directed edge always occurs between nodes belonging to different classes.

Bipartite graphs introduce the idea of capturing the relations among the members of a single class due to the interaction with the members of the other class. In fact, such a graph can be projected onto the predators overlap network, where two predators are connected with an edge weighted proportionally to the numbers of prey they have in common. Correspondingly, it can be projected into the prey overlap network, where two prey are connected according to the number of predators they share.

The projection graphs are two undirected, weighted graphs whose sizes are the number of possible predators $T+I$ and the number of possible prey $B+I$ in the food web respectively. 
The corresponding adjacency matrices $A^{pred}$ and $A^{prey}$ are symmetric and we define their elements in the following way:
\begin{eqnarray}
a_{ij}^{pred}=\frac{\sum_{k \in B+I} a_{ki} a_{kj} }{S(B+I)} \; with \; i,j \in T+I \label{pred}\\
a_{ij}^{prey}=\frac{\sum_{k \in T+I} a_{ik} a_{jk}}{S(T+I)} \; with \; i,j \in B+I  \label{prey}
\end{eqnarray}
We choose to normalize the predators weights over all possible prey, and the prey weights over all possible predators. Note that  $A^{pred}$ and $A^{prey}$ represent undirected graphs as  $a_{ij}^{pred}=a_{ji}^{pred}$ and $a_{ij}^{pred}=a_{ji}^{pred}$. Without loss of generality we can relabel the vertices in the two graphs keeping the information about the species they represent.

\begin{figure}[th]
\centerline{\psfig{file=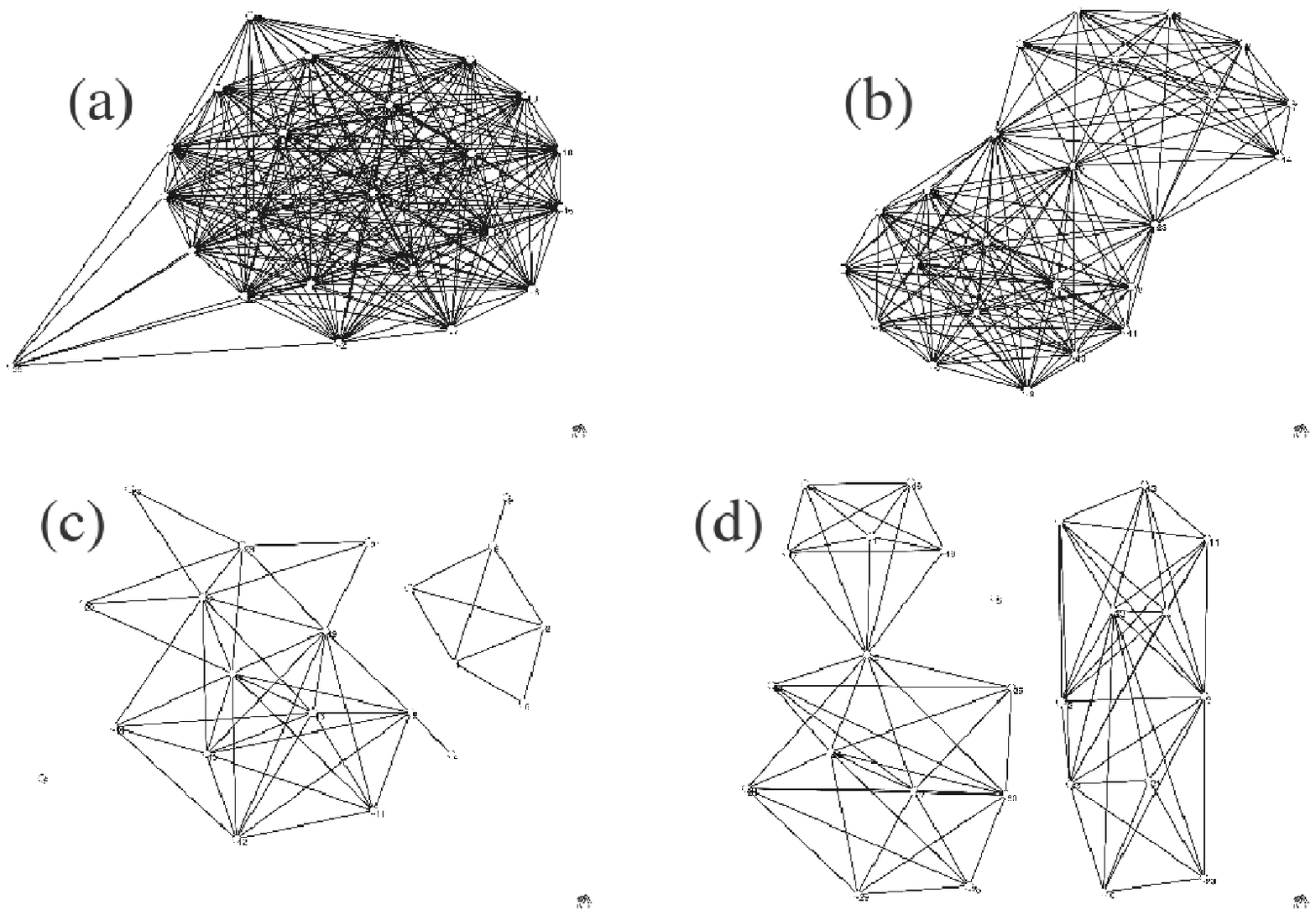,width=10cm}}
\vspace*{8pt}
\caption{Example of food-webs projection graphs: prey-prey from Skipwith Pond food web (a), predator-predator from Skipwith Pond food web (b), prey-prey from Chesapeake Bay food web (c), predator-predator from Chesapeake Bay food web (d).
}\label{fig:skipwith}
\end{figure}

Notice that the projection operation extends the ecological concept of niche overlap graphs \cite{nicheovergraph} that connect species insisting on the same niche; their combinatorial properties have been studied by \cite{suguihara}. Niche overlap graphs have been widely used in ecological literature as a measure of both dynamical \cite{abrams} and topological \cite{intervality} properties of food webs. Niche overlap graphs are therefore the unweighted version of projection graphs i.e., they can be obtained putting a link between each species connected by a non-zero weight in projection graphs. 

Example of projection graphs are given in Fig. \ref{fig:skipwith}. Looking at the empirical graphs we find a very symmetric topology (Fig. \ref{fig:skipwith}(a), \ref{fig:skipwith}(b)). Some projections show the formation of isolated communities both in predators and in prey graphs (Fig. \ref{fig:skipwith}(c), \ref{fig:skipwith}(d)) and sometimes we find only isolated nodes i.e., specialists, either in predation or in being a prey. Such community detection is a novelty compared to the classical studies of directed food webs: in fact, in both empirical and model graphs, there are no isolated nodes or clusters while in the niche model isolated species are removed.

\subsection{Topological Properties of Projection Graphs}

The weights of the projections graphs are a measure of the inter-specific competition for resources, giving information on how two species compete (or are objects of competition) in the predation interaction. A more significant meaning of this quantity should be derived by the analysis of an original weighted food web where the strength of the predation is also considered. In order to compare how much model webs reproduce inter-specific competition, we have measured and confronted the topological properties of projections graphs for both empirical and synthetic food webs.

The data we used have been selected to be the largest and highest-quality empirical trophic food webs present in literature. They represent a wide range of ecosystems, from freshwater habitat (Skipwith Pond SWP, Little Rock Lake LRL, Bridge Brooke Lake BBL) to freshwater-marine interface (Chesapeake Bay CPB, Ythan Estuary YE) to terrestrial habitats (Coachella Valley CDE, Saint Martin Island SMI)

We first calculate the path-length matrix whose elements are given by
\begin{equation}
d_{ij}^{pred/prey}=\min 
\lbrace 
	\sum_{k,l\in P_{ij}} \theta
			\left( a_{kl}^{pred/prey}\right)  
\rbrace  
\label{paths}
\end{equation}
where $P_{ij}$ is a path connecting node $i$ and $j$ and $\theta\left(x\right)$ is $1$ for $x>0$ and $0$ otherwise. We put $d_{ij}$ to $0$ when the nodes $i$ and $j$ belong to different clusters (i.e. no paths exists among them). 

A first characterization of the graph is given by the diameter $D$ corresponding to the maximum path length $d_{ij}$ occurring in the graph. Another characteristic length of the graph is the average path length, that can be computed using two different normalizations:
\begin{equation}
l_G=\frac{1}{n(n-1)/2}\sum_{i>j}d_{ij}\qquad l_R=\frac{1}{[n(n-1)/2]-(l_0/2)}\sum_{i>j}d_{ij} \label{averagelgt}
\end{equation}
where $l_0$ is the number of zeros in the matrix of path lengths. With these definitions $l_G$ is the average path length over all possible paths and $l_R$ is the average path length over all occurring paths. We notice that the average path length ratio $l_G/l_R$ is an indicator of the presence of separate components in the graph; in fact, $0\leq l_G/l_R \leq 1$ which is $0$ when all nodes are isolated and $1$ when all nodes are in the same cluster. 

We also characterize the local structure of the graphs by measuring the average clustering coefficient $Cl$. Despite the name, $Cl$ is not related to the presence of separate clusters but is a measure of how dense ("clustered") the graph is around a node: in fact, it is $1$ if all the neighbouring nodes are interconnected, and $0$ if there are no links between them.

A crucial quantity in many graphs is the degree of a node, i.e. the  number of its links \cite{CaldarelliBook}; in the case of weighted graphs, its natural extension is the  weight $w$ of a node defined as the sum of the weights of its links. We measure for our graphs the average weight $\langle w \rangle$.

Notice that, except for average weights, the measures of topological features we adopt would be the same for weighted and unweighted graphs; most of our conclusions would therefore apply also to niche overlap graphs. 

For real food webs we notice the formation of communities in both projections. Connectance varies from $0.24$ of CDE to $0.92$ of SWP, indicating that projections graphs are strongly connected in comparison to the original food webs. The average weights $\langle w \rangle$ take high values independent of the original connectance, varying from $0.04$ to $0.21$ for predators and from $0.05$ to $0.22$ for prey. Furthermore we notice that projections of food webs present all the characteristics of small world networks i.e. small diameter and large clustering \cite{smallworld}.

\begin{figure}[th]
\centerline{\psfig{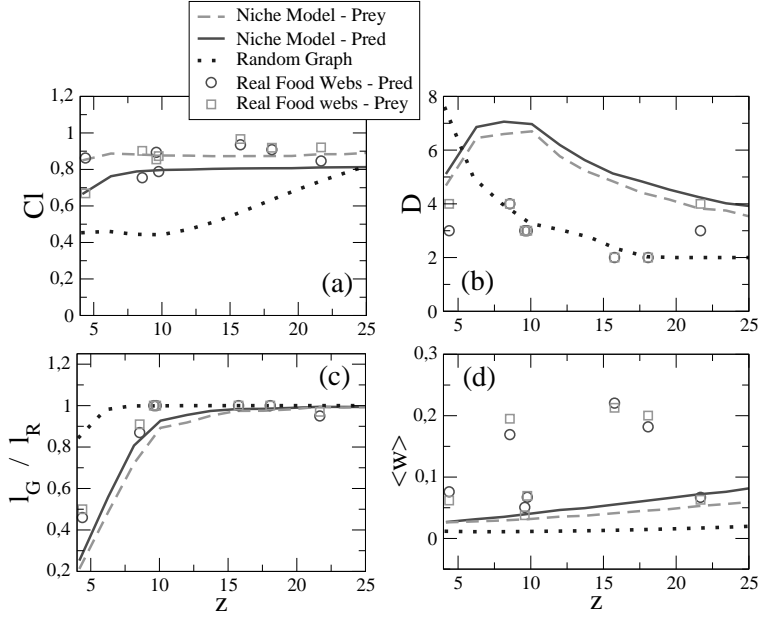}}
\vspace*{8pt}
\caption{Analysis of the topological features of projection graphs in function of the average degree of the original food web ($z$): clustering coefficient $Cl$ (a), diameter $D$ (b), the average path length ratio  $l_G/l_R$ indicating the presence of a single cluster (c), average weight $\langle w \rangle$ (d).
}\label{fig:topology}
\end{figure}

We then repeat the same analysis for the projections of synthetic webs derived from the niche model and the random graph model. Notice that the averaged topological properties of random projections are the same for predators' and prey's graphs in other words, in and out degree distributions of random digraphs have the same form \cite{newmanrand}. Comparing the curves with empirical data, we find good agreement for the clustering $Cl$ and average path lengths ratio $l_G/l_R$ (Fig. \ref{fig:topology}(a),(c)). Although the niche model represents the formation of competitive communities in empirical food webs better than the random graph model, it overestimates diameters and underestimates average weights (Fig. \ref{fig:topology}(b),(d)). We notice that empirical $D$ is better described by random graphs (Fig. \ref{fig:topology}(b)). 

From Fig. \ref{fig:topology}(d) we see how $\langle w \rangle$ for empirical food webs is larger than the values predicted by niche model or random graphs. We can explain this trend considering the feeding rule of niche model which assigns prey from a single portion of niche space. In doing so, the niche model has a reduced probability of sharing different resources (sharing resources is to share a single, well defined line interval): in this way the resulting graph has fewer paths, augmenting the diameter and reducing the weights of the projections graphs. On the other hand, empirical food webs are not strictly intervalled and do exhibit a strong bias towards contiguity of prey \cite{intervality}. This result suggests that empirically observed niches, once mapped onto a single dimension, could be composed of multiple intervals along niche space.

\section{Population Dynamics}

Given the network structure, we want to define the population dynamics for the individuals of the species described in this food web. We therefore associate with each node $i$ a population, i.e. a function of time $N_{i}(t)$ which represents the density of individuals of the same trophic species per unit of area.

To describe population dynamics we use the generalized Lotka-Volterra equations :
\begin{equation}
  \frac{dN_i}{dt} = r_i N_i \left( \frac{K_i - \sum_{j=1}^S \alpha_{ij}N_j}{K_i} \right) \label{lveq}
\end{equation}
where $r_{i}$ is the intrinsic growth rate of species $i$, $K_{i}$ his carrying capacity and $\alpha_{ij}$ represents the effect species $j$ has on the population of species $i$. Pulling the carrying capacity into the interaction term the equations became
\begin{equation}
    \frac{dN_i}{dt} = r_i N_i \left( 1 - \sum_{j=1}^S \overline{\alpha_{ij}}N_j \right) \label{lvgeneral}
\end{equation}
This doesn't actually change the equations, but only how the interaction $\overline{\alpha_{ij}}=\alpha_{ij}/K_i$ is defined. For simplicity all self-interacting terms $\overline{\alpha_{ii}}$ are set to 1.

One can represent both the populations and the growth rates as rows of numbers (vectors) and the interaction term $\alpha$ as a matrix, called also \emph{competition kernel}. Let us suppose that we have only one type of external resource $R$ produced with a constant rate $y$ (\emph{renewability} of resources) and let us also suppose that each basal species consumes a fraction $X$ at a rate $c_{i}$. The equation for the resources is:
\begin{equation}
  {dR(t) \over dt} = R(y-\sum_{i=1}^Sc_iN_i) \label{res}
\end{equation}
where the first term represents the renewal rate and the second one gives the total rate of consumption\footnote{here is assumed that $c_{i}\neq 0$ only for basal species.}.

If we consider the equilibrium conditions for equations that is \ref{lveq} and \ref{res}, $dN_i/dt=0$ and $dR/dt=0$, we find a relationship between the ecological parameters given by
\begin{equation}
y=\sum_{ij}^Sc_i\alpha_{ij}^{-1}K_j \label{constraint}
\end{equation}
This gives a fundamental constraint on all the parameters (except the $r_i$), especially on the competition kernel which must be invertible.

Some authors use as competition kernel a function of the distances between species in niche space \cite{kernel,gaussian}. The problem with such a choice is that the topology of the food web is not included in the equations; a simple way to incorporate it is to use a combination of projections weights. For this reason, we consider equation (\ref{lveq}) and propose the following competition kernel that joins in a self-organized way\cite{self} the topology with the dynamics:
\begin{equation}
\alpha_{ij}= a_{ij}^{pred}-a_{ij}^{prey} \label{projkernel}
\end{equation}
This means that the influence of population $i$ on population $j$ is
negative if species $i$ and $j$ share some prey and positive if they share some predators. The competition between two species increases with respect to  the number of prey species they share and vice versa.  Using the elements defined in eq. \ref{projkernel}, 
it is possible to simulate a population dynamics on both model and empirical food webs.

The stability of empirical and model food webs has been tested  numerically following this steps:

\begin{enumerate}
\item \emph{Food web adjacency matrix}. We generate it using niche model, where the input parameters are connectance $C$ and size $S$. In real cases this is done considering empirical food webs data.

\item \emph{Competition kernel}. 
We derive the competition kernel $\alpha_{ij}(C)$ from equation \ref{projkernel}.

\item \emph{Ecological parameters}. We fix the parameters $r_i$, $c_i$ and $K_i$:

\begin{itemize}
\item As a first approximation the intrinsic growths $r_i$ are set equal from all species. This means that in the ideal condition of no competition and infinite resources, all the populations should grow at the same rate. Varying these parameters one can simulate the different lifetimes of species and their reproduction strategies.

\item Setting the carrying capacities $K_i$ of the species means that, in the ideal condition of no inter-specific competition ($\alpha_{ij}=0$), the maximum number of individuals per unit of space which are sustainable from the external environment is fixed for every species.

\item the consuming rates $c_i$ has been set to 0.1 only for basal species. This is quite realistic because basal species are, by definition, the species which directly feed on the external environment.

\item At this point using equation \ref{constraint} we fix the renewability of resources $y$ which depends basically on the topology of the graph. To avoid indefinite growth, for simulated food webs, we fix the maximum renewability $y_M=10$ and generate graphs with the same connectance and size until we obtain the desired renewability ($0<y\leq y_M$).
\end{itemize}

\item \emph{Integration of equations}. Once we have the competition kernel and the desired $y$ we solve the equations \ref{lveq} using the 4th order Runge-Kutta algorithm

\item \emph{Stability}. The results have been derived in the steady state. 
\end{enumerate}

\subsection{Dynamical properties}
We test the stability of these Lotka-Volterra systems using both empirical and model food webs. We remark that using the coefficients \ref{projkernel} as the competition kernel, the system quite always reaches the steady state for small $S$ and $C$. For large values of $S$ and $C$, the system is mostly unstable and populations sizes go to infinity when the complexity  $S\,C$ grows. We tested the robustness of the steady states changing both initial conditions and ecological parameters.

When the intrinsic growth rates are augmented the steady state is reached faster. Population sizes at the steady state should be equal to the carrying  capacities, but the presence of competition kernel changes the effective $K_i$ of every species. However, this parameters give the ordering of sizes of stable populations and when $K_i=K_P$ the mean population is always $K_p$.

\begin{figure}[th]
\centerline{\psfig{file=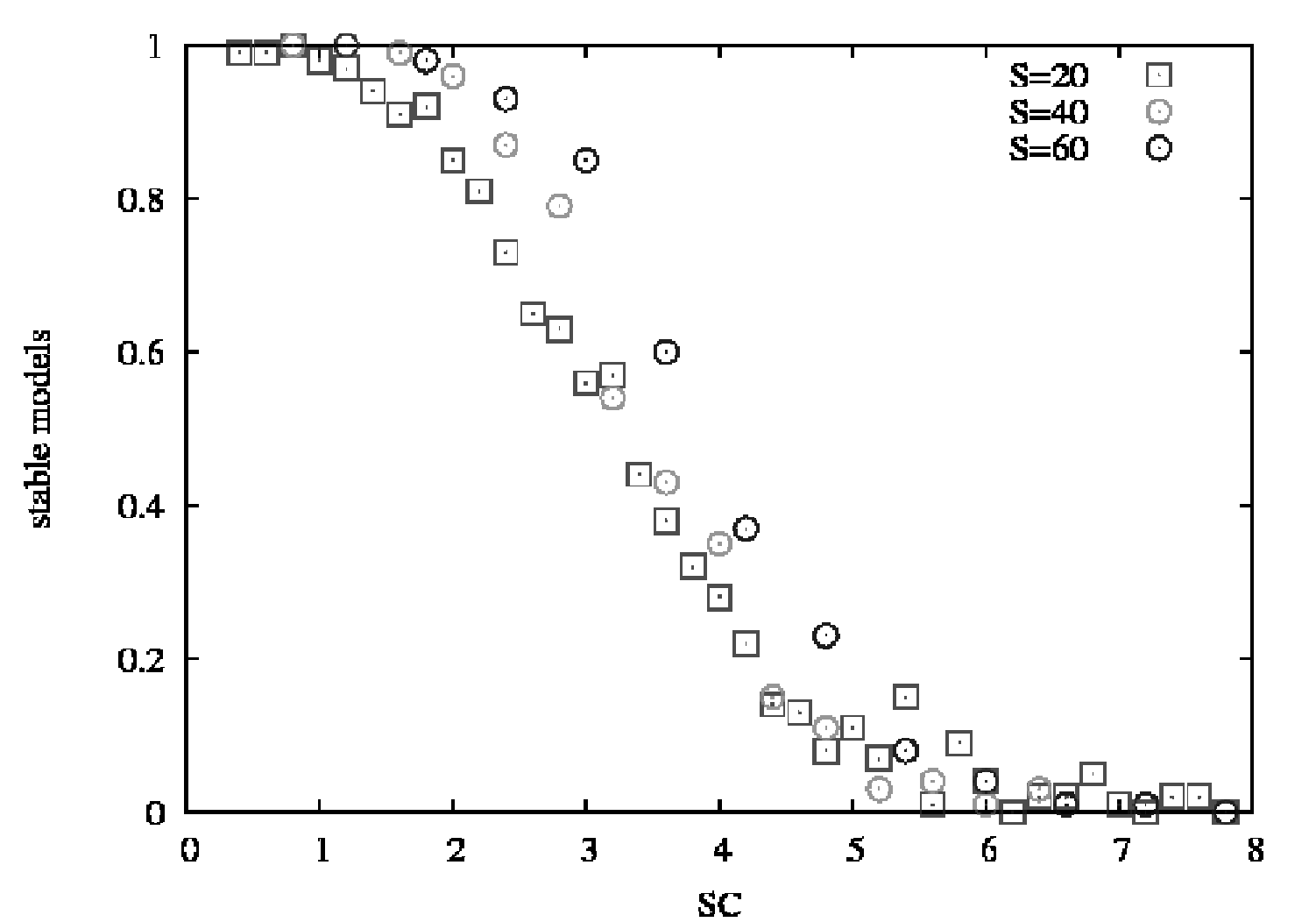,width=10cm}}
\vspace*{8pt}
\caption{Dynamical stability expressed as the fraction of non diverging models versus complexity.
}\label{fig:csdyn}
\end{figure}

With our choice for the kernel, the only stable empirical food web is Chesapeake Bay (CPB) which is the one with the lowest ecological complexity $SC$. To investigate the effects of the complexity $SC$ on the stability, we simulate and analyse population dynamics on the niche model. We generate $100$ realizations of synthetic graphs from the niche models for  fixed values of $S$ and $C$; on such graph, we test the stability of population dynamics starting from random and uniform initial conditions. We find that, using expression \ref{projkernel} as competition kernel, the only two observed behaviours are reaching a steady state (stability) or divergence; we count the number of times each of these behaviours occurs. As expected from general consideration of the dynamical stability of population dynamics \cite{LordMay}, the probability of reaching a steady state decreases with complexity (Fig. \ref{fig:csdyn}).

We find that population dynamics on graphs from the niche model is stable only for small $S$ and $C$ ; the model exhibits stability for $SC<2$ with stability that decreases linearly between $SC=2$ and $SC=5$. These results are robust under the change of initial conditions and ecological parameters.

\section{Conclusions}

We have introduced weighted projection graphs as a tool to analyse prey-prey and predator-predator relations. 

Building on the standard representation of food webs as directed networks, projection graphs extend  the concept of the niche overlap graphs by adding the possibility of having weighted links. Projection graphs for ecological data reveal community structures both among preys and predators; in general, they can be used to detect prey-prey or predator-predator indirect effects. 

We analyse the projection graphs derived both from ecological data and from two synthetic ecological models: a simple null case (random graph) and a successful synthetic food-web generator (the niche model). The comparison of the topological features of projection graphs derived from synthetic data and real data indicates some direction of improvement for the niche model; in particular, the largeness of the diameter of projection graphs for the niche models indicates the necessity either to extend the niche model to multiple intervals or to consider a multi-dimensional representation of niches.

We further investigate the Lotka-Volterra population dynamics under the assumption that the competition kernel can be inferred from the prey-prey and predator-predator interactions described by the weights of the projected graphs.
Analysing the behaviour of the dynamics for niche model food webs of increasing complexity, we find that according to May's classical results for the stability criterion, the probability that a food web's dynamics reaches a stable state decreases with its complexity, measured as the number of interactions among network's elements.

\section*{Acknowledgments}
GM would like to thank Daniel Stouffer, Andrew Beckerman and Paulo Guimaraes for interesting discussion about competition and niche overlap graphs and an anonimous reviewer for very helpful remarks.
We thank captain Francesco Totti and admiral Agostino Straulino for strong psycological support and nursing. Even if it is hard to believe, the funding of the project is due to the italian government.

\vspace*{-3pt}   

\bibliographystyle{ws-acs}



\end{document}